\documentclass[showpacs,reprint,amsmath,amssymb,aps]{revtex4-1}
\usepackage{}
\usepackage{cases}
\usepackage{amsmath}
\usepackage{amssymb}
\usepackage{amsfonts}
\usepackage{amssymb}
\usepackage{dcolumn}
\usepackage{bm}
\usepackage{hyperref}
\usepackage{graphicx}
\usepackage{upgreek}

\begin{document}
\title{Coherence-protected Quantum Gate by Continuous Dynamical Decoupling in Diamond}
\author{Xiangkun Xu}
\author{Zixiang Wang}
\author{Changkui Duan}
\author{Pu Huang}
\author{Pengfei Wang}
\author{Ya Wang }
\author{Nanyang Xu}
\author{Xi Kong}
\author{Fazhan Shi}
\author{Xing Rong}
\author{Jiangfeng Du}
\altaffiliation{djf@ustc.edu.cn}
\affiliation{Hefei National Laboratory for Physics Sciences at
Microscale and Department of Modern Physics, University of Science
and Technology of China, Hefei, 230026, China}

\begin{abstract}

To implement reliable quantum information processing, quantum gates have to be protected together with the qubits from decoherence. Here we demonstrate experimentally on nitrogen-vacancy system that by using continuous wave dynamical decoupling method, not only the coherence time is prolonged by about 20 times, but also the quantum gates is protected for the duration of controlling time. This protocol shares the merits of retaining the superiority of prolonging the coherence time and at the same time easily combining with quantum logic tasks. It is expected to be useful in task where duration of quantum controlling exceeds far beyond the dephasing time.

\end{abstract}


\maketitle

Decoherence of the quantum system is one of the main obstacles for the realization of quantum information processing, quantum simulation and quantum sensing \cite{Nielson-Chuang,Ladd-Nature2010}. Quantum gates as primary elements for large-scale universal quantum information processing are unavoidably affected by decoherence, and have to be implemented within the decoherence time. Therefore a qubit needs to be protected not only when it is idle but also during the process of quantum gate operations. Among various approaches of quantum coherence protection, dynamical decoupling (DD) \cite{Viola} is a particularly promising strategy for its integratability with quantum gates by elaborate designs.

Protecting quantum logic tasks with pulsed DD has been proposed \cite{Linda,rbliu,Uhrig,dassarma,west} and experimentally demonstrated in a few quantum systems for extending the coherence of individual qubit \cite{nature2009,nature-bollinger,Science-DD,cory,suter} and two qubit entanglement \cite{wangya}. Despite the great success of pulsed DD in protecting the coherence, it becomes complicate when combing with quantum gates by considering the commutation between the decoupling pulses and the control pulses which is by no means trivial.
This can be seen from the recent demonstration on single NV center by adapting the time intervals between the dynamical decoupling pulses \cite{nature-Hanson}.

In order to retain the flexibility of carrying out quantum logic tasks while the qubits are protected, the continuous wave dynamical decoupling (CWDD) approach has been developed \cite{Lukin-theory,pra,cai,Bermudez,gong}. The application of CWDD to quantum gate protection has only been implemented very recently to trapped ions \cite{nature-trapped ion}. In this letter, we present the combination of CWDD with quantum gates to the solid-state system of NV center
 in diamond. By applying continuous microwave driving fields the decoherence is suppressed and coherent time is extended by more than an order of magnitude, and by encoding the qubit in microwave dressed state the performance of quantum gate has been protected far beyond the quantum system's free induced decay (FID) time at room temperature. we show that while retaining the superiority of prolonging the coherence time, protected quantum logic tasks can be implemented in an almost trivial way.

NV center system is deliberately chosen in this study for its potential in solid-state quantum computing. NV centre spin qubits are promising for quantum information processing due to fast resonant spin manipulation \cite{giga}, long coherence time \cite{nature-material}, easy initialization and read-out by laser \cite{first}. However, the long coherence time of NV center is not immediately exploitable \cite{nature-material}, one has to decouple the NV center from the magnetic interactions with its spin based environment \cite{Science-DD,cory}. To fully exploit the merits of NV centers for quantum information processing, decoupling NV centers from the unwanted magnetic interaction with their spin-based environment while at the same time implementing desired quantum gate operations is essential.

\begin{figure}[htbp]
\centering
\includegraphics[width=1\columnwidth]{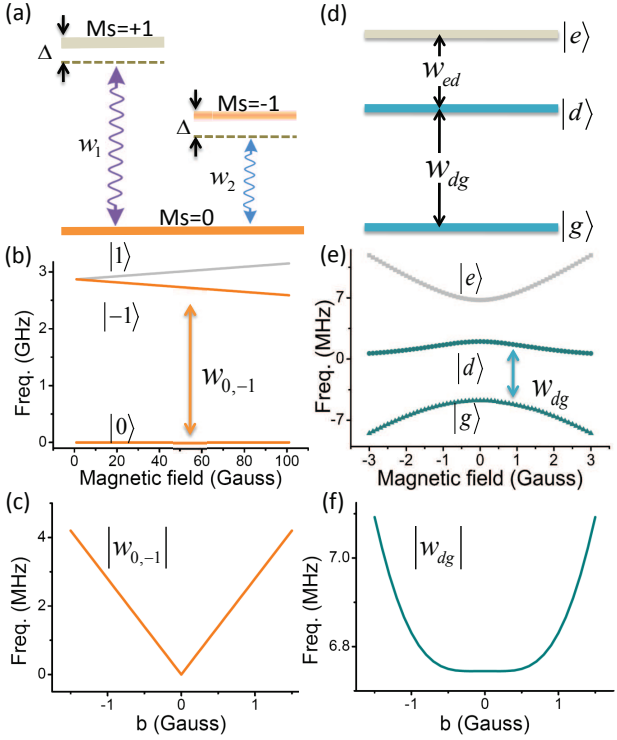}
  \caption{(Color online).
   (a) The electron spin-triplet states (thick solid lines) of NV Center in an external magnetic field $B_z$ along the symmetry axis ([111] direction). The purple and blue wavy arrow lines indicate two applied continuous microwaves on frequency $w_{1}$ and $w_{2}$ with same amplitude, corresponding to the same Rabi frequency $\Omega$.
   (b) The splitting of the electron spin-triplet states of NV Center with the external magnetic field $B_z$, with $w_{0,-1}$ the energy gap between states $|-1\rangle$ and $|0\rangle$ representing the qubit; (c) The sensitivity of the $w_{0,-1}$ to fluctuation filed b;
   (d) The dressed-states of the driven NV Center system of (a), with $w_{dg}$ and $w_{eg}$ showing the corresponding energy gaps; (e) The dependence of the energies of dressed states and (f) the sensitivity of the $w_{dg}$ to the fluctuation filed $b$, parameters used: detuning $\Delta=2$ MHz and Rabi frequency $\Omega=4\Delta$ for both microwaves.
}
  \label{fig1}
\end{figure}

 \begin{figure}[htbp]
\centering
\includegraphics[width=1\columnwidth]{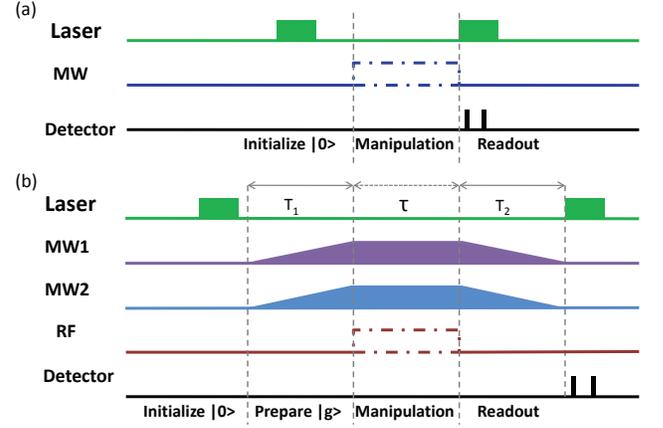}
  \caption{(Color online).
   Diagram of the experimental pulse sequences used in the experiment. (a) Pulse sequences without CWDD. The electron spin state of NV center is initialized to $|M_{s}=0\rangle $ by the 532-nm laser pulse, manipulated by microwave pulses, and read out through the fluorescence. (b) Pulse sequences with CWDD. The 532 nm laser initializes the electron spin state of NV center to $|M_{s}=0 \rangle$. Then two microwave driving amplitude of field (MW1,MW2) are ramped linearly to time $T_1$, the ground state $|g\rangle$ in the driven space is adiabatically prepared. Then the driving amplitudes are kept constant, during this time an additional Radio-Frequency(RF) is applied to realize manipulation. Finally the adiabatical process transfer the microwave dressed state in the protected subspace to the bare electronic spin states of the NV center and readout by the fluorescence.}
  \label{fig2}
\end{figure}

Here is the theory for the protection of the qubit and the quantum logic for a single NV center using CWDD method.
The electronic spin ground states of NV center in an external field $B_z$ along the symmetry axis can be described by the following Hamiltonian:
\begin{equation}
\label{eq£º1}
H = DS_{z}^{2}+\gamma_eB_zS_z.
\end{equation}
The zero-field splitting with $D=2.87$ GHz and the Zeeman term with $\gamma_e=2.802$ MHz/Gauss determine the eigenstates $|M_s\rangle$ ($M_s=\pm 1,\ 0$) shown in Fig.\ 1(a) by thick lines. The loss of quantum coherence of NV center in Type IIa diamond is mainly caused by surrounding $^{13}\rm C$ nuclear spin bath fluctuations \cite{huangpu,zhaonan}. This can be described by an effective random magnetic field
$b$ with time correlation \cite{Science-DD}.
The linear dependence of the energies $|\pm 1\rangle$ on magnetic field (Fig. 1(b)) results into sharp variation of the energy gap $w_{0,-1}$ (Fig. 1(c)) between the two states of the qubit ($|0\rangle$ and $|-1\rangle$). This results into strong decoherence due to the random magnetic field $b$.

An equal-weighted superposition of $|\pm 1\rangle$ can result into a state with eigenvalue insensitive to the random field $b$. This can be realized by applying off-resonant continuous microwave driving fields at the same time, as shown in Fig.\ 1(a) by $w_1$ and $w_2$. The Hamiltonian of the NV center driven by two microwaves of the same off-resonance $\Delta$ and Rabi frequency $\Omega$ can be written in the interaction picture as:
\begin{equation}
\label{eq£º2}
H_{\rm NV}=\sum_{\imath=\pm 1}\left ((\Delta+\gamma_e b \cdot \imath)|\imath\rangle\langle\imath|+\frac{\Omega}{2}\left (|0\rangle\langle\imath|+|\imath\rangle\langle0|\right ) \right )
\end{equation}
The diagonalization of $H_{\rm NV}$ results into three dressed-states $|e\rangle$, $|d\rangle$ and $|g\rangle$ in the driven space, as depicted in Fig. 1(d).
The two lower levels $|g\rangle$ and $|d\rangle$ of a gap $w_{dg}$ is used as a qubit in the driven space. Due to the symmetry of $H_{\rm NV}$ and nonzero $\Omega$, the energies of all the three states depend only on ${b}^2$ (shown in Fig.\ 1(e)) in contract to the linear dependence of the bare states shown in Fig. 1(b), the dephasing between $|d\rangle$ and $|g\rangle$ is strongly suppressed when the microwave driving amplitude $\Omega$ is much larger than the effective random field. With appropriate ratio of $\Omega/\Delta$ ($=4$) shown in Fig. 1(f), even the $ {b}^2$ term in $w_{dg}$ is eliminated so that the lowest order is $\sim {b}^4$ and much greater coherence time can be achieved in the protected subspace spanned by $|d\rangle$ and $|g\rangle$ \cite{Lukin-theory}. The gradual transfer from $|0\rangle $ to $|g\rangle$ with the MW Rabi frequency $\Omega$ and the nonzero $S_z$ between $|g\rangle $ and $|d\rangle$ allow adiabatical preparation and readout, and RF manipulation of the qubit in the driven space.

The whole experimental scheme for characterizing the NV center and quantum logic operations for the two cases without or with CWDD is given in Fig. 2.
In the undriven NV system the experimental pulses are depicted in Fig. 2(a). The laser pulses are used for initial state preparation and final state readout, and the microwave (MW) pulses are used to manipulate the qubit. While in the driven system (Fig.2(b)), after the initialization of $|0\rangle$ two driven microwaves (MW1,MW2) are up-ramped linearly to prepare $|g\rangle$ adiabatically. Then the RF is used for the qubit manipulation encoded in the driven system while with two microwave continuous protection. At last, the down-ramped microwaves maps the encoded state $|g\rangle$ ($|d\rangle$) to $|0\rangle$ ($|-1\rangle$) for laser readout.

In this experiment, the sample used is type IIa (with nitrogen density $<5$ ppb). A 12 Gauss magnetic field generated by three pairs of Helmholtz coils was used to remove the degeneration of $M_{s}=\pm 1$ states. The magnetic field is aligned with the NV symmetry axis. The two protecting microwave fields are generated by the sidebands of a local oscillator (LO) mixed with an Integrated Frequency (IF) which produced by a Direct Digital Synthesizer (DDS). The detuning $\Omega$ is controlled by LO. The phase and amplitude of the microwaves (MW1, MW2) used in Fig. 2(b) can be controlled by the DDS. Then through a linear amplifier (16 W) the microwaves are radiated to the NV center via a $20\ \mu$m copper wire with the 50 Ohm resist terminator. The controlled RF is provided with a 10 MHz Arbitrary Waveform Generator (AWG) and directly coupled to the copper line. All input signals are synchronized by a pulse generator. In the experiment, the length of initialization laser pulse is $3\ \mu$s and the waiting time following the laser is $5\ \mu$s. The photoluminescence (PL) is measured during an integration time of $0.35\ \mu$s. To suppress the photon statistic error, each measurement is typically repeated more than $10^6$ times.

 The optical detected magnetic resonance spectrum for $|0\rangle \longrightarrow |1\rangle$ transition in undriven NV system is shown in Fig. 3(a). The splitting is caused by hyperfine coupling with $^{14}\rm N$ nuclear spin. Here and below we only consider the subspace $M_{I}=0$ of $^{14}\rm N$ nuclear spin. The error of resonate frequency $w_{\pm 1}$ of the $|0 \rangle \leftrightarrow|\pm1 \rangle$ in the subspace of $M_{I}=0$ is within $\pm 20$ KHz. In undriven NV system, the FID signal oscillates and dephases on a fast time scale, as shown in Fig. 3(b).
 The oscillation of FID is caused by the beats originated from three transitions of different frequencies as a result of hyperfine splitting related to the nuclear spin $I=1$ of the $^{14}\rm N$.
 The damping of the oscillation of FID signal is well fitted with a Gaussian envelope function $\exp[-(t/T_{2}^{*})^{2}]$ \cite{Science-DD}, giving a dephasing time $T_{2}^{*} = 0.93\ \mu$s.

\begin{figure}[htbp]
\centering
\includegraphics[width=1\columnwidth]{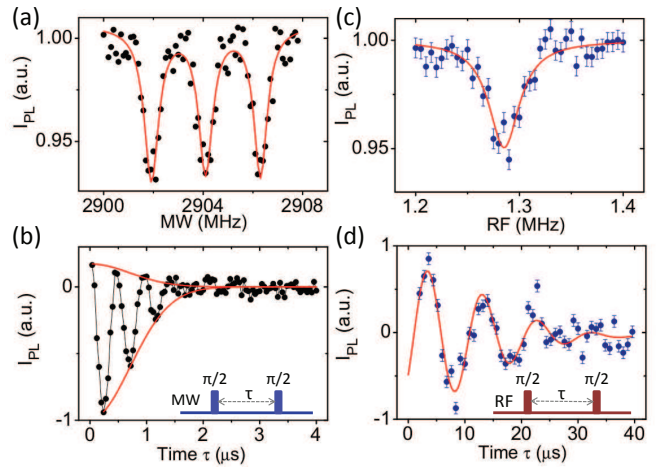}
  \caption{(Color online).
  (a) The ODMR spectrum for transition between $|0\rangle$ to $|1\rangle$ in NV center (black dots).The smooth pink curve is plot to guide eyes. The three lines due to hyperfine coupling with $^{14}\rm N$ nuclear spin are resolved.
  (b)Measured (black dots) and fitted (red line) free induced decay (FID) signals of the NV center. The oscillations and the asymmetric envelopes are due to coupling to the $^{14}\rm N$ nuclear spin. The MW sequences in the lower right corner is used in the manipulation part in Fig. 2(a).
  (c)The ODMR spectrum for continuous microwave driven NV center system. The dip in the figure corresponds to the transition between state $|d\rangle$ and $|g\rangle$. Error bars are standard deviation.
  (d) Experimental free induced decay (FID) signals in the NV center driven system. The RF sequence in the lower right corner is used in the manipulation part in Fig. 2b.}
  \label{fig3}
\end{figure}

To implement quantum logic in the driven space, the resonant R.F. frequency $w_{dg}$, i.e., the energy gap between $|g\rangle$ and $|e\rangle$ needs to be experimentally determined first. With the experimental parameters for the driving MW fields: $\Omega = 1.6$ MHz and $\Delta = 0.4$ MHz (i.e., the frequencies of MW1 and MW2 are $w_{0,1}-\Delta$ and $w_{0,-1}-\Delta$) and forward and reversed ramp time $T_1 = T_2 = 50\ \mu$s, the $w_{dg}$ is determined by sweeping the RF frequency using a duration of the RF pulse of 20 $\mu$s.
The result is shown in Fig. 3(c) and it shows a much shaper dip. The FID signal in the subspace protected by CWDD is shown in Fig. 3(d). The oscillation in the FID signal is due to the deviation of the resonant position in the protected subspace. The data is fitted to $\exp[-(t/T_{2,\rm CWDD}^{*})^{2}]\cos (2 \pi f t + \varphi)$ where $f$ value corresponds to the deviation. The value of $T_{2,\rm CWDD}^{*}$ is derived to be $18.9\ \mu$s from the fitting. This shows that the coherence time of the CWDD protected single spin is prolonged from the coherence time of the bare spin (of $0.93 \ \mu$s) of NV center by about 20 times.

\begin{figure}[htbp]
\centering
\includegraphics[width=1\columnwidth]{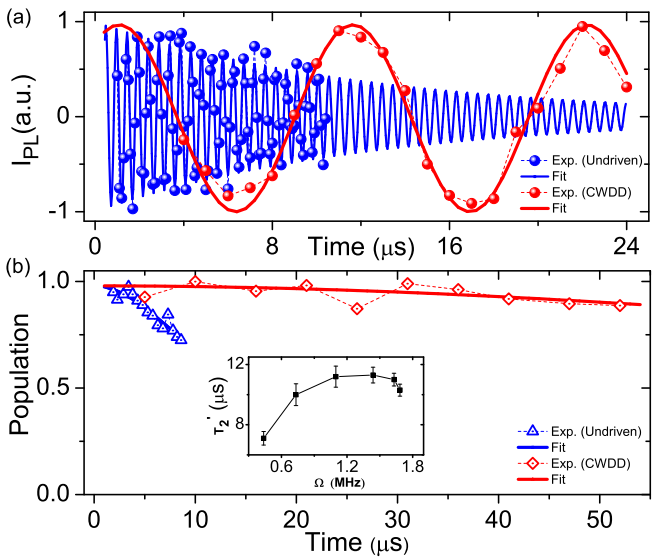}
  \caption{(Color online).
   (a) Measured decay of the Rabi oscillations both in the undriven space (blue dots) and in the subspace protected by CWDD(red dots), the curve is for guiding eyes only. (b) The population of the designated state as a result of the successive NOT gate operation in undriven space with MW $\uppi$ pulse (blue triangles), or in CWDD protected space with RF $\uppi$ pulse (red diamonds). The curve is fitted to Gauss lineshape decays. Subset is the measured decay time $T_2^{\prime}$ as a function of the microwave field amplitude $\Omega$.}

  \label{fig4}
\end{figure}

The manipulations of the unprotected qubit using MW pulses (scheme in Fig. 2(a)) and the protected qubit using RF pulses (scheme in Fig. 2(b)) are carried out to show the performance of the CWDD approach.

Fig. 4(a) plots the Rabi oscillations in the $M_{s}=0,-1$ basis without driven (blue dots) and {$|e\rangle$, $|g\rangle$} with CWDD (red dots). It clearly shows that the oscillation in the {$|e\rangle$, $|g\rangle$} is well-preserved almost without decay even after 25 $\mu$s, while that in $M_{s}=0,-1$ basis suffers considerable decay even in a few $\mu$s. Due to the complexity of measuring the fidelity of the quantum gate in the protected space, the performance of the CWDD protected quantum gate is evaluated using the coherence of the state after successive NOT gate operations. The performance of the NOT gate is shown in Fig. 4(b) by plotting the indicator $F$ defined to be $|\langle \psi | d\rangle|^{2}$ for odd number of NOT gates and $\langle \psi | g\rangle|^{2}$ for even number of NOT gates. Compared to the decay time of 11 $\mu$s in the undriven case, for the CWDD driven case, the coherence is maintained far beyond 50 $\mu$s, clearly demonstrating the performance of the quantum gate protected with CWDD.

To further understand the decay of the coherence in the Fig. 4, we plot in the subset of Fig. 4(b) the measured $T_2^{\prime}$ of the bare state of the NV center as a function of the amplitude of the microwave fields (of frequencies $\omega_{0,1}$ and $\omega_{0,-1}$ but the same amplitude). It clearly shows that the $T_2^{\prime}$ increases first and then decreases fast with increasing microwave strength. This is attributed to the effect of microwave field fluctuation, similar to previous experiments { \cite{quantum beats}}. When the microwave field amplitude is small, the fluctuation amplitude is also small and the main contribution to dephase is due to the restrained bath fluctuation, which leads to an increase of $T_2^{\prime}$. As the MW field amplitude increases further, the fluctuation becomes larger and eventually its effect exceeds that of the bath fluctuation. This leads to sharp decrease of $T_2^{\prime}$. It is expected that in the case of the microwave field fluctuation being suppressed, much longer $T_2^{\prime}$ could be achieved. It is noted that our results shown in Fig. 4 have already revealed that the applied RF field suppresses the MW field fluctuation.

In summary, we have realized CWDD in the solid-state system of NV centers in diamond. The coherence time of NV center is prolonged by about $20$ times with CWDD. What is more important, we have combined the CWDD with quantum gate operation. The performance of quantum gate is greatly improved compared to the same quantum manipulation without CWDD. Although being demonstrated in diamond here, the method can also be applied to other systems, such as other ion-doped crystals and quantum dots etc.

This work was supported by National Nature Science
Foundation of China (Grants Nos.10834005, 91021005
and 11161160553), the Instrument Developing Project of
the Chinese Academy of Sciences(Grant No. Y2010025),
and the National Fundamental Research Program.

\end{document}